# A mathematical structure for the generalization of the conventional algebra

A. El Kaabouchi[†], L. Nivanen[†], Q. A. Wang[†], J. P. Badiali[‡], and A. Le Méhauté[†]

[†]Institut Supérieur des Matériaux et Mécaniques Avancés du Mans
44 Avenue Bartholdi, 72000 Le Mans, France

[‡]UMR 7575 LECA ENSCP-UPMC,
11 rue P. et M. Curie, 75231 Cedex 05, Paris, France

**Abstract**

An abstract mathematical framework is presented in this paper as a unification of several deformed or generalized algebra proposed recently in the context of generalized statistical theories intended to treat certain complex thermodynamic or statistical systems. It is shown that, from mathematical point of view, any bijective function can be used in principle to formulate an algebra in which the conventional algebraic rules are generalized.

# 1 Introduction

In the last two decades, statistical physics has experienced several extensions from the conventional Boltzmann-Gibbs-Shannon (BGS) formalism to other seemingly more general formalisms. Among these extensions, there are two ones called the nonextensive statistical mechanics (NSM) [1][2] and the $\kappa$-statistical mechanics (KSM)[3]. Both of them *overcome*, among others, the limitative rule of additivity of energy and entropy, a paradigm governing the conventional statistical mechanics. Due to the strong relationship between non extensive statistical physics and fractal geometry [4] it has been claimed that these statistics can be used to describe complex systems whose anomalous behaviors cannot be interpreted within BGS statistics. Since the proposition of these extensions, much pros and cons has been written from the physical point of view. We will not enter into the physical debate here. The present work is limited in the pure mathematical study of some algebraic aspects related to the mathematical functionals used in the extended statistics.

The first functional have been used in NSM and called $q-$exponential $e_a^x = (1+ax)^{1/a}$ and $q-$logarithm $\ln_a x = \dfrac{x^a - 1}{a}$ where $a = 1-q$ or $a = q-1$. They are inverse functions of one another. When $q$ tends toward 1, or when $a$ tends toward 0, the $q-$exponential tends toward the usual exponential function and the $q-$logarithm tends toward the usual logarithm. From mathematical point of view, these functions have interesting properties which extend partially those of the corresponding usual functions. For example, the usual exponential and logarithm entails the usual algebraic operators such as addition, subtraction, multiplication and division. In mimicking the related morphisms, an extended algebra has been defined from $e_a^x$ and $\ln_a x$ using generalized operators [5][6]. Further development shows application in the derivation of $q-$factorial and $q-$multinomial coefficient [7].

The second functional called $\kappa-$exponential are used in KSM. In that case $\exp_{\{\kappa\}}(x) = \left(\kappa x + \sqrt{1+\kappa^2 x^2}\right)^{1/\kappa}$ and $\kappa-$logarithm $\ln_\kappa x = \dfrac{x^\kappa - x^{-\kappa}}{2\kappa}$. They are also inverse function of one another. In the case of $\kappa$ tends toward 0, these generalized functions recover the usual ones. An extended $\kappa-$algebra is also developed from these functions [3].

Other deformations or extensions of conventional statistics with extended algebra [8][9] were also proposed. A common character of extended algebra is to use a generalization of exponential function or its inverse to expand the morphism of, say, a usual algebraic operator […]. For example, using $e_a^x = (1+ax)^{1/a}$, if we calculate the usual product of two exponential, we find $e_a^x \times e_a^y = e_a^{x+_a y}$ where the operator $+_a$ is the generalized $q$-addition given by $x +_a y = x + y + axy$.

The present work is an extension of this logic within a more general mathematical framework. It will be shown that the same methodology can be carried out at least with any bijection, not only with the generalized exponentials. The concomitant mathematical structure can yield different algebraic rules according to the choice of different functions and different field of their functional definition. We think that this formulation of the extended algebra may be beneficial for further development and understanding of the new mathematical tendency stimulated by the development of physics.

## 2  Structure of generalized group and ring

### 2.1 Preliminaries

Let $X$ and $Y$ be two nonempty sets, and let $*$ and $\perp$ be two binary operations on $X$ and $Y$ respectively. We denoted by $1_X$ and $1_Y$ the neutral element (when this exists) of $(X,*)$ and $(Y,\perp)$ respectively, and by $x^{-1}$ the inverse of $x$ if $x$ is invertible in $(X,*)$ or in $(Y,\perp)$. We mean by $\varphi$ a bijection of $(X,*)$ in $(Y,\perp)$ and by $\psi$ the inverse function of $\varphi$.

### 2.2 Definitions and theorem

a) The functions $+_{(\varphi,\perp)}$ and $-_{(\varphi,\perp)}$ defined on $X \times X$ by :

$$\forall (x,x') \in X \times X, \; x +_{(\varphi,\perp)} x' = \psi(\varphi(x) \perp \varphi(x')).$$

$$\forall (x,x') \in X \times X, \; x -_{(\varphi,\perp)} x' = \psi\left(\varphi(x) \perp (\varphi(x'))^{-1}\right) \quad \text{(when this exists)}$$

are binary operations (laws) on $X$ ;

b) The functions $\times_{(f,*)}$ and $\div_{(f,*)}$ defined on $Y \times Y$ by

$$\forall (y,y') \in Y \times Y, \; y \times_{(\varphi,*)} y' = \varphi(\psi(y) * \psi(y')).$$

$$\forall (y,y') \in Y \times Y, \; y \div_{(\varphi,*)} y' = \varphi\left(\psi(y) * (\psi(y'))^{-1}\right) \quad \text{(when this exists)}$$

are binary operations (laws) on $Y$ ;

c) $\varphi$ is an homomorphism of $(X,+_{(\varphi,\perp)})$ in $(Y,\perp)$, and $\psi$ is an homomorphism of $(Y,\times_{(\varphi,*)})$ in $(X,*)$ ;

d) When $\varphi$ is an homomorphism of $(X,*)$ in $(Y,\perp)$ then $+_{(\varphi,\perp)} = *$ and $\times_{(\varphi,*)} = \perp$. In particular, $\varphi$ becomes an isomorphism of $(X,+_{(\varphi,\perp)})$ into $(Y,\times_{(\varphi,*)})$.

## 2. 3 Examples

a) When $(X,*) = (\mathbf{R},+)$, $(Y,\perp) = (\mathbf{R}_+^*,\times)$ and $\varphi = \exp$, then $\psi = \ln$.

We have for all $(x,x') \in X \times X$ and $(y,y') \in Y \times Y$

$$x +_{(\varphi,\perp)} x' = \psi(\varphi(x) \times \varphi(x')) = \ln(\exp x \times \exp x') = x + x'$$

$$x -_{(\varphi,\perp)} x' = \psi(\varphi(x) \times (\varphi(x'))^{-1}) = \ln\left(\frac{\exp x}{\exp x'}\right) = x - x'$$

$$y \times_{(\varphi,*)} y' = \varphi(\psi(y) + \psi(y')) = \exp(\ln y + \ln y') = y \times y'$$

$$y \div_{(\varphi,*)} y' = \varphi(\psi(y) - \psi(y')) = \exp(\ln y - \ln y') = y \div y'.$$

we find thus the ordinary operations.

b) Let $a \in \mathbf{R}_+^*$ and we consider the set $X = \left]-\frac{1}{a},+\infty\right[$ with the law additive, here $* = +$), and $(Y,\perp) = (\mathbf{R}_+^*,\times)$. We mean by $\varphi$ the function of $X$ in $(Y,\perp)$ defined in [4] [5] by

$$\varphi(x) = (1+ax)^{1/a} = e_a^x.$$

It's easy to check that $\varphi$ is a bijection of $X$ into $Y$, and that the inverse function of $\varphi$ is defined by

$$\forall x \in Y, \; \psi(x) = \ln_a x = \frac{x^a - 1}{a}.$$

In this case, we have for all $(x,x') \in X \times X$ and $(y,y') \in Y \times Y$ according to generalized operations [5][6]

$$x +_{(\varphi,\perp)} x' = \psi(\varphi(x) \times \varphi(x')) = x + x' + axx' = x +_a x'$$

$$x -_{(\varphi,\perp)} x' = \psi(\varphi(x) \times (\varphi(x'))^{-1}) = \frac{x - x'}{1 + ax'} = x -_a x'$$

$$y \times_{(\varphi,*)} y' = \varphi(\psi(y) + \psi(y')) = (y^a + y'^a - 1)^{1/a} = y \times_a y'$$

$$y \div_{(\varphi,*)} y' = \varphi(\psi(y) + (\psi(y'))^{-1}) = (y^a - y'^a - 1)^{1/a} = y \div_a y'.$$

c) Let $\kappa \in \mathbf{R}_+^*$ and we consider $(X,*)=(\mathbf{R},+)$, $(Y,\perp)=(\mathbf{R}_+^*,\times)$. Let $\varphi$ the function of $X$ in $(Y,\perp)$ defined by

$$\forall x \in X, \ \varphi(x) = \left(\kappa x + \sqrt{1+\kappa^2 x^2}\right)^{1/\kappa} = \exp_{\{\kappa\}}(x).$$

$\varphi$ is bijective of $X$ into $Y$, and that the inverse function of $\varphi$ is defined by

$$\forall x \in Y, \ \psi(x) = \ln_{\{\kappa\}}(x) = \frac{x^\kappa - x^{-\kappa}}{2\kappa}.$$

In this case, we have all $(x,x') \in X \times X$ and $(y,y') \in Y \times Y$

$$x +_{(\varphi,\perp)} x' = \psi(\varphi(x) \times \varphi(x')) = x\sqrt{1+\kappa^2 x^2} + x'\sqrt{1+\kappa^2 x'^2} = x \overset{\kappa}{\oplus} x'$$

$$x -_{(\varphi,\perp)} x' = \psi(\varphi(x) \times (\varphi(x'))^{-1}) = x\sqrt{1+\kappa^2 x^2} - x'\sqrt{1+\kappa^2 x'^2} = x \overset{\kappa}{\ominus} x'$$

$$y \times_{(\varphi,*)} y' = \varphi(\psi(y)+\psi(y')) = \frac{1}{\kappa}\operatorname{sh}(\operatorname{argsh}(\kappa y) + \operatorname{argsh}(\kappa y')) = y \overset{\kappa}{\otimes} y'$$

we find thus the operations introduced by Kaniadakis [3][8].

d) Let $\kappa \in \mathbf{R}_+^*$ and we consider $r \in [-\kappa, \kappa]$, $(X,*)=(\mathbf{R},+)$, $(Y,\perp)=(\mathbf{R}_+^*,\times)$. Let $\psi$ the function of $Y$ in $(X,*)$ defined by

$$\forall x \in Y, \ \psi(x) = \ln_{\{\kappa,r\}}(x) = \frac{x^{r+\kappa} - x^{r-\kappa}}{2\kappa}.$$

We check easily that $\psi$ is bijective of $Y$ into $X$ ($\psi$ is continuous and strictly increasing on $Y$). We denote by $\varphi : x \mapsto \varphi(x) = e_{\{\kappa,r\}}(x)$ the inverse function of $\psi$.

We note that $\psi$ verifies the following property

$$\forall y, y' \in Y, \ \psi(yy') = y^{\kappa+r}\psi(y') + y'^{\kappa+r}\psi(y) - 2\kappa\psi(y)\psi(y').$$

Its follows that for all $(x,x') \in X \times X$, we have

$$x +_{(\varphi,\perp)} x' = \psi(\varphi(x) \times \varphi(x')) = x \overset{\kappa,r}{\oplus} x'.$$

See [8].

e) Let $\gamma \in \mathbf{R}_+^*$ and we consider $\kappa = 3\frac{\gamma}{2}$, $r = \frac{\gamma}{2}$, $(X,*)=(\mathbf{R},+)$, $(Y,\perp)=(\mathbf{R}_+^*,\times)$. Let $\psi$ the function of $Y$ in $(X,*)$ defined by

$$\forall x \in Y, \ \psi(x) = \ln_\gamma(x) = \ln_{\{\kappa,r\}}(x) = \frac{x^{r+\kappa} - x^{r-\kappa}}{2\kappa} = \frac{x^{2\gamma} - x^{-\gamma}}{3\gamma}.$$

We denote by $\varphi : x \mapsto \varphi(x) = e_\gamma(x)$ the inverse function of $\psi$, then $\varphi$ is defined by

$$\forall x \in X, \ \varphi(x) = e_\gamma(x) = \left[ \left( \frac{1 + \sqrt{1 - 4\gamma^3 x^3}}{2} \right)^{1/3} + \left( \frac{1 - \sqrt{1 - 4\gamma^3 x^3}}{2} \right)^{1/3} \right]^{1/\gamma}$$

We note that $\psi$ verifies the following property

$$\forall y, y' \in Y, \ \psi(yy') = y\psi(y') + y'\psi(y) - 3\gamma\psi(y)\psi(y').$$

Its follows that for all $(x, x') \in X \times X$, we have

$$x +_{(\varphi, \perp)} x' = \psi(\varphi(x) \times \varphi(x')) = x \overset{\gamma}{\oplus} x'.$$

See [8].

f) Let $\gamma \in \mathbf{R}_+^*$ and we consider $(X,*) = (\mathbf{R},+)$, $(Y,\perp) = (]0,1],\times)$. Let $\varphi$ the function of $(X,+)$ into $(Y,\perp)$ defined in [10][11][12] by

$$\forall x \in X, \ \varphi(x) = \exp(-x^\gamma)$$

The inverse function $\psi$ of $\varphi$ is defined by

$$\forall x \in Y, \ \psi(x) = \left( \ln\left(\frac{1}{x}\right) \right)^{\frac{1}{\gamma}}$$

We check easily that for all $(x, x') \in X \times X$ and $(y, y') \in Y \times Y$, we have

$$x +_{(\varphi, \perp)} x' = \psi(\varphi(x) \times \varphi(x')) = \left( x^\gamma + x'^\gamma \right)^{\frac{1}{\gamma}}$$

$$y \times_{(\varphi, *)} y' = \exp\left( -\left[ \left( \ln\left(\frac{1}{y}\right) \right)^{\frac{1}{\gamma}} + \left( \ln\left(\frac{1}{y'}\right) \right)^{\frac{1}{\gamma}} \right]^\gamma \right).$$

As an example, see the following operation for $\gamma = 2$:

$$\forall (x, x') \in \mathbf{R}_+^* \times \mathbf{R}_+^*, \ x +_{(\varphi, \perp)} x' = \sqrt{x^2 + x'^2}.$$

We observe that the Gauss distribution is related to Pythagoras relation on the circle. This observation is a key observation to fully understand the relation between diffusion process and 2D fractal Dynamics [4].

**2.4 Remarks**

a) Let $a > 0$. We show that $\varphi : x \mapsto \varphi(x) = e_a^x$ is the *unique* function defined on $X = \left] -\frac{1}{a}, +\infty \right[$, differentiable at 0 with $\varphi'(0) = 1$ and verifies for all $(x, x') \in X \times X$

$$\varphi(x +_a x') = \varphi(x) \times \varphi(x').$$

b) Let $\kappa > 0$. We show that $\varphi : x \mapsto \varphi(x) = \exp_{\{\kappa\}}(x)$ is the *unique* function defined on $\mathbf{R}$, differentiable at 0 with $\varphi'(0) = 1$ and verifies for all $(x, x') \in X \times X$

$$\varphi(x \overset{\kappa}{\oplus} x') = \varphi(x) \times \varphi(x').$$

**2.5 Proposition** With the previous notations, we have

$$+_{(\varphi, \times_{(\varphi,*)})} = *  \quad \text{and} \quad \times_{(\varphi, +_{(\varphi,\perp)})} = \perp.$$

**2.6 Proposition** Let $x, x', x'' \in X$, we have the following properties

$$\varphi\big(x * (x' +_{(\varphi,\perp)} x'')\big) = \varphi(x) \times_{(\varphi,*)} \big(\varphi(x') \perp \varphi(x'')\big)$$

$$\varphi\big((x' +_{(\varphi,\perp)} x'') * x\big) = \big(\varphi(x') \perp \varphi(x'')\big) \times_{(\varphi,*)} \varphi(x).$$

**2.7 Remarks**

a) The commutativity of the law $*$ (respectively $\perp$) implies the commutativity of the law $\times_{(\varphi,*)}$ (respectively $+_{(\varphi,\perp)}$).

b) The associativity of the law $*$ (respectively $\perp$) implies the associativity of the law $\times_{(\varphi,*)}$ (respectively $+_{(\varphi,\perp)}$).

c) The existence of a neutral element of $(X, *)$ (respectively of $(Y, \perp)$), implies the existence of a neutral element $(Y, \times_{(\varphi,*)})$ (respectively of $(X, +_{(\varphi,\perp)})$).

d) Due to the fact that the two laws are not defined in the same set there is no reason to consider the distributivity of $\times_{(\varphi,*)}$ with respect to $+_{(\varphi,\perp)}$ or conversely.

We can therefore make the following proposition.

**2.8 Proposition**

With the previous notations, we have

a) $(Y, \perp)$ is an abelian group if and only if $(X, +_{(\varphi,\perp)})$ is an abelian group ;

b) $(X, *)$ is an abelian group if and only if $(Y, \times_{(\varphi,*)})$ is an abelian group ;

c) $(Y, \perp, \circ)$ is a ring if and only if $(X, +_{(\varphi,\perp)}, +_{(\varphi,\circ)})$ is a ring ;

d) $(Y, \perp, \times_{(\varphi,*)})$ is a ring if and only if $(X, +_{(\varphi,\perp)}, *)$ is a ring.

## 3. Structure of generalized vector space

**3.1 Definition and theorem.** Let $X$ and $Y$ be two nonempty sets, $\bullet$ a external law on $Y$, We mean by $\varphi$ a bijection of $X$ in $Y$ and by $\psi$ the inverse function of $\varphi$.

The function $\bullet_{(\varphi,\bullet)}$ defined on $\mathbf{K} \times X$ by

$$\forall (\lambda, x) \in \mathbf{R} \times X, \ \bullet_{(\varphi,\bullet)} (\lambda, x) = \psi(\lambda \bullet \varphi(x))$$

is a external law on $X$.

We denote $\bullet_{(\varphi,\bullet)} (\lambda, x)$ by $\lambda \bullet_{(\varphi,\bullet)} x$.

**3.2 Applications**

a) When $\mathbf{K} = \mathbf{R}$, $X = \mathbf{R}$, $Y = \mathbf{R}_+^*$ and $\bullet$ is the law defined on $Y$ by

$$\forall (\lambda, x) \in \mathbf{R} \times Y, \ \lambda \bullet x = x^\lambda.$$

If we consider the function $\varphi$ of $X$ in $Y$ defined by

$$\forall x \in X, \ \varphi(x) = \exp x.$$

Then for all $(\lambda, x) \in \mathbf{R} \times X$,

$$\lambda \bullet_{(\varphi,\bullet)} x = \ln(\lambda \bullet \varphi(x)) = \lambda x.$$

In this case we denote $\bullet_{(\exp,\bullet)}$ in stead of $\bullet_{(\varphi,\bullet)}$.

We find thus the ordinary multiplication.

b) Let $a \in \mathbf{R}_+^*$. When $X = \left]-\dfrac{1}{a}, +\infty\right[$, $\mathbf{K} = \mathbf{R}$, $Y = \mathbf{R}_+^*$ and $\bullet$ is the law defined on $Y$ by

$$\forall (\lambda, x) \in \mathbf{R} \times Y, \ \lambda \bullet x = x^\lambda.$$

If we consider the function $\varphi$ from $X$ in $Y$ defined by

$$\forall x \in X, \ \varphi(x) = (1+ax)^{1/a} = \mathrm{e}_a^x.$$

Then for all $(\lambda, x) \in \mathbf{R} \times X$,

$$\lambda \bullet_{(\varphi,\bullet)} x = \psi(\lambda \bullet \varphi(x)) = \dfrac{(1+ax)^\lambda - 1}{a}.$$

In this case we denote $\bullet_{(a,\bullet)}$ in stead of $\bullet_{(\varphi,\bullet)}$.

We find thus the operation defined in [4][5].

c) Let $\kappa \in \mathbf{R}_+^*$. When $\mathbf{K} = \mathbf{R}$, $X = \mathbf{R}$, $Y = \mathbf{R}_+^*$ and $\bullet$ is the law defined on $Y$ by

$$\forall (\lambda, x) \in \mathbf{R} \times Y, \ \lambda \bullet x = x^\lambda.$$

If we consider the function $\varphi$ from $X$ in $Y$ defined by

$$\forall x \in X, \; \varphi(x) = \left(\kappa x + \sqrt{1 + \kappa^2 x^2}\right)^{1/\kappa} = \exp_{\{\kappa\}}(x).$$

Then for all $(\lambda, x) \in \mathbf{R} \times X$,

$$\lambda \bullet_{(\varphi, \bullet)} x = \ln_{\{\kappa\}}(\lambda \bullet \varphi(x)) = \frac{\left(\kappa x + \sqrt{1 + \kappa^2 x^2}\right)^{\lambda} - \left(\kappa x + \sqrt{1 + \kappa^2 x^2}\right)^{-\lambda}}{2\kappa}$$

In this case we denote $\bullet_{(\{\kappa\}, \bullet)}$ in stead of $\bullet_{(\varphi, \bullet)}$.

### 3.3 Remarks

a) We show that $\varphi : x \mapsto \varphi(x) = \exp x$ is the *unique* function defined on $\mathbf{R}$, differentiable at 0 with $\varphi'(0) = 1$ and verifies

$$\forall (\lambda, x) \in \mathbf{R} \times \mathbf{R}, \; \varphi\left(\lambda \bullet_{(\exp, \bullet)} x\right) = (\varphi(x))^{\lambda}.$$

b) Let $a > 0$. We show that $\varphi : x \mapsto \varphi(x) = e_a^x$ is the *unique* function defined on $X = \left]-\frac{1}{a}, +\infty\right[$, differentiable at 0 with $\varphi'(0) = 1$ and verifies

$$\forall (\lambda, x) \in \mathbf{R} \times X, \; \varphi\left(\lambda \bullet_{(a, \bullet)} x\right) = (\varphi(x))^{\lambda}.$$

c) Let $\kappa > 0$ We show that $\varphi : x \mapsto \varphi(x) = \exp_{\{\kappa\}}(x)$ is the *unique* function defined on $\mathbf{R}$, differentiable at 0 with $\varphi'(0) = 1$ and verifies

$$\forall (\lambda, x) \in \mathbf{R} \times X, \; \varphi\left(\lambda \bullet_{(\{\kappa\}, \bullet)} x\right) = (\varphi(x))^{\lambda}.$$

### 3.4 Theorem

If $(Y, \perp, \bullet)$ is a vector space on $\mathbf{K}$, then $\left(X, +_{(\varphi, \perp)}, \bullet_{(\varphi, \bullet)}\right)$ is a vector space on $\mathbf{K}$.

## 4 Conclusion

This paper reportes a general framework which unifies all the extended algebra recently proposed from physical considerations. Our main conclusion is that (at least) any bijective function can be the characteristic function of an algebra which may generalize the conventional algebra characterized by the usual exponential function.

It is for algebraic and functional reasons that the exponential function play a major role both in mathematics and in physics. This role is essentially rightly related with elementary mathematical manipulations and the fitting with the behaviour of separable physical phenomena. Nevertheless the physics of complex media breaks this relevance in different way: nonextensivity, correlation and coupling, scaling properties etc [10]. In that case the exponential function and all related properties loss their physical accuracy and are transformed into power laws. The present mathematical analysis shows not only how this accuracy is lost but moreover what are the key factors which must rebuild up to handle again the complex problematic.

It must be observed that the generalisation of exponential analysis suggested in the present paper is based on the link between functional definitions and there algebraic and functional fields of validities. This link which is a key factor to understand the dynamics of physical phenomena in complex geometry is currently in progress. Another work relative to the generalisation of usual differential calculus with bijective and non bijective functions will be presented in the near future.